\documentclass[conference]{IEEEtran}

\usepackage{amsmath,amssymb,amsfonts}
\usepackage{graphicx}
\usepackage{booktabs}
\usepackage{siunitx}
\usepackage{mathtools}
\usepackage{bm}
\usepackage{xcolor}
\usepackage{url}
\usepackage{tikz}
\usepackage[hidelinks]{hyperref}
\usepackage{cleveref}

\newcommand{\ket}[1]{\left|#1\right\rangle}
\newcommand{\bra}[1]{\left\langle#1\right|}

\graphicspath{{figures/}{images/ieee/}}

\begin{document}

\title{Variational Joint Magnetometry and Gradiometry on Dipolar Spin Chains}

\author{
\IEEEauthorblockN{Priyam Srivastava\textsuperscript{1,\dag}, Xin Jin\textsuperscript{2}, Junyu Liu\textsuperscript{2}, Gurudev Dutt\textsuperscript{3},\\
Tom Purdy\textsuperscript{3}, Kang Kim\textsuperscript{4}, Kaushik P. Seshadreesan\textsuperscript{1}}
\IEEEauthorblockA{
\textsuperscript{1}\textit{Dept. of Informatics \& Networked Systems, University of Pittsburgh, Pittsburgh, PA, USA}\\
\textsuperscript{2}\textit{Dept. of Computer Science, University of Pittsburgh, Pittsburgh, PA, USA}\\
\textsuperscript{3}\textit{Dept. of Physics \& Astronomy, University of Pittsburgh, Pittsburgh, PA, USA}\\
\textsuperscript{4}\textit{Dept. of Bioengineering, University of Pittsburgh, Pittsburgh, PA, USA}\\[0.4ex]
Emails: \texttt{prs216@pitt.edu}\textsuperscript{\dag}, \texttt{XIJ90@pitt.edu}, \texttt{junyuliu@pitt.edu}, \texttt{gdutt@pitt.edu},\\ \texttt{tpp9@pitt.edu}, \texttt{kangkim@upitt.edu}, \texttt{kausesh@pitt.edu}\\[0.4ex]
\textsuperscript{\dag}Corresponding author
}
}

\maketitle

\begin{abstract}
Estimating a uniform magnetic field $B_0$ and its spatial gradient $g$ on a dipolar-coupled spin chain calls for a multiparameter figure of merit. The GHZ state, optimal for single-parameter Heisenberg-limited sensing, has a rank-one quantum Fisher information matrix with $\det(Q^{\mathrm{GHZ}}) = 0$ at every chain length $N$, ruling it out for the two-parameter problem. We present a variational framework that takes $\det(F)$ as the objective and a hardware-motivated layered dipolar circuit as the ansatz. Both encoding generators are diagonal in the computational basis, which reduces the search for the quantum Fisher information benchmark to a probability-simplex optimization and yields a tractable best-found benchmark $\det(Q^*)$ against which variational performance is compared. The same diagonal structure makes the classical Fisher information depend only on basis-state probabilities under any single-qubit decoder, so encoder and decoder parameters are co-trained with CMA-ES in a single run. Decoder optimization past fixed Ramsey adds at most a few percentage points across the grid, in contrast to the persistent decoder gains seen in our prior single-parameter work. Variational probes at $L = 3$ reach $0.92$ of the best-found benchmark at $N = 5$, a $4.2\times$ SQL advantage in $\det(F)$, and concentrate on a four-string motif of the two GHZ extrema and two half-chain-flip strings whose structure follows from the Dicke-sector decomposition of the two generators.
\end{abstract}


\section{Introduction}
\label{sec:intro}

Quantum sensors that share quantum correlations between spatially distributed probes can attain precisions beyond the limits of independent probes~\cite{lorenzo_2006,Giovannetti2011}. The ultimate precision attainable by a probe state is bounded by its quantum Fisher information, while the precision realized by a specific measurement is captured by the classical Fisher information~\cite{Liu2020-pk}. For separable probes, the classical Fisher information scales linearly with the number of probes $N$ and defines the standard quantum limit (SQL). Entangled probes can achieve quadratic scaling, the Heisenberg limit, with the GHZ state as the optimal probe under uniform single-parameter encoding and standard Ramsey readout.

Many practical sensing tasks are not single-parameter. A representative example arises in nanoscale magnetometry on dipolar-coupled spin systems, where nitrogen-vacancy (NV) centers in diamond have become a leading platform for wide-field imaging of magnetic fields from biological and geological samples~\cite{degen2017,Rondin2014-hm,barry2020}. The signal of interest in these settings is the spatial gradient of the field, but it sits on top of a bias $B_{0}$ that is rarely stable on the integration timescale. Coil-current fluctuations, temperature-dependent zero-field-splitting shifts, and stray fields all contribute to drift~\cite{Pham2011-qu,Maletinsky2012}, so a bias value calibrated at the start of an experiment becomes uncertain by the end, and that uncertainty enters the gradient estimate the moment $B_{0}$ is treated as known. Estimating $B_{0}$ and $g = \partial B / \partial x$ together from the same measurement record sidesteps the calibration-drift floor and replaces it with the intrinsic Cram\'er-Rao bound on the pair.

Quantifying that bound requires moving from a scalar to a matrix description. The classical Fisher information matrix $F$ carries diagonal elements that report individual sensitivities to $B_{0}$ and $g$, together with an off-diagonal element that measures the statistical correlation between the two estimators~\cite{Liu2020-pk}. The natural scalar summary is the determinant $\det(F)$, the information volume, which penalizes both small diagonal sensitivities and large cross-correlations at once. This is the D-optimal design criterion from classical experimental design~\cite{Kiefer1959}, and the appropriate choice when both parameters are co-equal estimation targets. The choice has concrete consequences for what quantum advantage is even possible. The standard nuisance-parameter framing of Apellaniz~\textit{et al.}~\cite{gradiometry_toth} reports no gradient advantage over the SQL on the equidistant chain, while the same chain admits a best-found benchmark of roughly $4.5\times$ the SQL at $N = 5$ once $\det(F)$ is adopted as the objective.

A direct application of single-parameter intuition fails here. The GHZ state, which saturates the entanglement-enhanced bound for uniform single-parameter encoding, gives a rank-one quantum Fisher information matrix with $\det(Q^{\mathrm{GHZ}}) = 0$ exactly for every $N$. The state is sensitive only to a single linear combination of $B_{0}$ and $g$ and is blind to the orthogonal direction in parameter space, so its quantum advantage in single-parameter sensitivity does not transfer. No closed-form expression is known for the optimal probe state on the equidistant chain, and variational optimization over a hardware-motivated ansatz~\cite{Meyer2021-zl,PhysRevLett.120.080501} becomes a constructive necessity for this problem.

\subsection*{Main contributions and findings}

In this work we present a variational framework for joint $(B_{0}, g)$ estimation on a linear chain of dipolar-interacting spin-1/2 systems, building on the variational multiparameter toolbox of Ref.~\cite{Meyer2021-zl} and the layered dipolar ansatz developed in our prior single-parameter work~\cite{srivastava2024pra}. Our contributions are:

\begin{enumerate}
\item \textbf{Explicit GHZ collapse on the equidistant chain.} We show that the GHZ state, optimal for single-parameter Heisenberg-limited sensing, has rank-one QFIM and therefore $\det(Q^{\mathrm{GHZ}}) = 0$ at every $N$ on the joint $(B_0, g)$ problem. The collapse motivates variational design as a constructive necessity rather than a convenience and is the analytical anchor for the sector-structure interpretation of the optimized state in Sec.~\ref{sec:results:discussion}.

\item \textbf{A simplex reduction for the quantum bound.} We exploit the diagonality of both encoding generators in the computational basis to reduce the search for the quantum upper bound from the $2^{N}$-dimensional Hilbert space to the $(2^{N}-1)$-dimensional probability simplex, yielding a tractable numerical benchmark $\det(Q^{*})$ against which variational performance is benchmarked.

\item \textbf{A universal four-string motif in the optimal probe.} We identify a four-string motif, consisting of the two GHZ extrema and two half-chain-flip strings, in the structure of the variationally optimized probe state across all $(L, N)$ cells of the optimization grid, and trace its emergence to the Dicke-sector structure of the two generators.

\item \textbf{Joint encoder--decoder optimization with a near-redundant decoder.} We jointly optimize encoder and decoder parameters across a four-tier decoder hierarchy in a single CMA-ES run, and find that decoder optimization past fixed Ramsey contributes at most a few percentage points to the achievable $\det(F)$. This stands in contrast to the structured single-parameter setting of our prior work~\cite{srivastava2024pra}, where decoder gains under non-uniform encoding persist at depth.
\end{enumerate}

Numerically, variational probes reach $0.92$ of $\det(Q^{*})$ at $(L = 3, N = 5)$ under fixed Ramsey readout, realizing $4.2\times$ the SQL at the same cell. The depth cap $L \leq 3$ reflects the near-term hardware regime, where compounding errors from entangling operations confine reliable state preparation to shallow circuits. This proof-of-principle study is performed under noiseless evolution to isolate the multiparameter structure of the problem from decoherence effects.

The remainder of the paper is organized as follows. Section~\ref{sec:setup} introduces the linear chain geometry and the phase encoding. Section~\ref{sec:fisher} defines the classical Fisher information matrix and the parameter-shift evaluation pipeline. Section~\ref{sec:bounds} derives the standard quantum limit and the GHZ collapse and describes the numerical computation of $\det(Q^{*})$. Section~\ref{sec:variational} specifies the variational ansatz, the four decoder tiers, and the CMA-ES optimization protocol. Section~\ref{sec:results} presents the scaling results and the state-structure analysis. Section~\ref{sec:conclusion} summarizes the findings and outlines extensions.

%
%
%
\section{Physical Setup}
\label{sec:setup}

\subsection{Linear Spin Chain and Field Model}
\label{sec:setup:geometry}

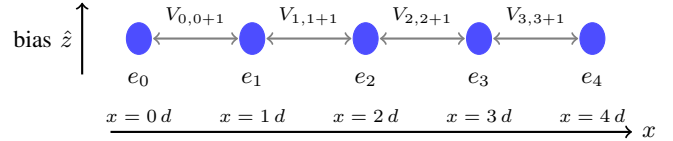
\begin{figure}[htbp]
  \centering
  \begin{tikzpicture}[xscale=0.75, yscale=0.95]
    \foreach \i in {0,1,2,3,4} {
      \filldraw[blue!70] (\i*2, 0) circle (0.22);
      \node[below] at (\i*2, -0.35) {\small $e_{\i}$};
      \node[below] at (\i*2, -0.85) {\scriptsize $x = \i\,d$};
    }
    \foreach \i in {0,1,2,3} {
      \draw[thick, gray, <->] (\i*2+0.25, 0) -- (\i*2+1.75, 0);
      \node[above] at (\i*2+1.0, 0.05) {\scriptsize $V_{\i,\i+1}$};
    }
    \draw[->, thick] (-1.0, -0.5) -- (-1.0, 0.5);
    \node[left] at (-1.0, 0) {\small bias $\hat{z}$};
    \draw[->, thick] (-0.5, -1.3) -- (8.7, -1.3) node[right] {\small $x$};
  \end{tikzpicture}
  \caption{Linear chain geometry for $N = 5$ dipolar-coupled spin-$\tfrac{1}{2}$ systems. Spin qubits at positions $x_i = i\,d$ are connected by dipolar couplings $V_{ij}$ that enter the entangling Hamiltonian (Sec.~\ref{sec:variational}).}
  \label{fig:geometry}
\end{figure}

We consider $N$ dipolar-coupled spin-$\tfrac{1}{2}$ systems arranged in a one-dimensional chain along the $\hat{x}$ axis~\cite{gradiometry_toth}. The chain is the simplest geometry that admits a position-weighted gradient generator while remaining hardware-realistic for NV-center electron-spin arrays~\cite{Maletinsky2012}, where each spin is implemented as the two-level subspace $\{\ket{0}, \ket{-1}\}$ of the NV ground-state triplet, addressed by resonant microwave control under a moderate bias field along $\hat{z}$ that lifts the $\ket{\pm 1}$ degeneracy. Spin $i$ is located at position
\begin{equation}
  x_i \;=\; i\,d, \qquad i = 0, 1, \dots, N-1,
  \label{eq:positions}
\end{equation}
where $d$ is the inter-spin spacing. The chain geometry and the dipolar couplings $V_{ij}$ are illustrated in Fig.~\ref{fig:geometry}, with the entangling Hamiltonian itself defined in Sec.~\ref{sec:variational}.

The chain is exposed to a one-dimensional magnetic field profile
\begin{equation}
  B(x_i) \;=\; B_0 \,+\, g\,x_i,
  \label{eq:field_model}
\end{equation}
with the unknown parameter vector
\begin{equation}
  \mathbf{x} \;=\; (B_0,\, g)^{\top} \;\in\; \mathbb{R}^{2}.
  \label{eq:param_vector}
\end{equation}
The linear expansion in Eq.~\eqref{eq:field_model} is valid when the array length $(N-1)\,d$ is small compared to the field variation length scale, which holds across the parameter range studied here. We work in natural units throughout, with $\gamma_e t = 1$ and $d = 1$.\footnote{Fisher information values are recovered by reinstating these factors. Derivatives with respect to $B_0$ acquire a factor $\gamma_e t$, while derivatives with respect to $g$ acquire a factor $\gamma_e t\,d$. Thus $F_{B_0 B_0}$, $F_{B_0 g}$, and $F_{g g}$ acquire factors $(\gamma_e t)^2$, $(\gamma_e t)^2 d$, and $(\gamma_e t)^2 d^2$, respectively, and $\det(F)$ acquires $(\gamma_e t)^4 d^2$ overall.}

\subsection{Phase Encoding and Generators}
\label{sec:setup:encoding}

Under the field model of Sec.~\ref{sec:setup:geometry}, spin $i$ accumulates a phase
\begin{equation}
  \phi_i \;=\; B_0 \,+\, g\,(i\,d),
  \label{eq:phase_per_spin}
\end{equation}
linear in both unknown parameters. The phase vector $\boldsymbol{\phi} = (\phi_0, \dots, \phi_{N-1})^{\top}$ is related to $\mathbf{x}$ by
\begin{equation}
  \boldsymbol{\phi} \;=\; M\,\mathbf{x}, \qquad
  M \;=\; \begin{pmatrix}
    1 & 0 \\
    1 & d \\
    1 & 2\,d \\
    \vdots & \vdots \\
    1 & (N-1)\,d
  \end{pmatrix},
  \label{eq:sensing_matrix}
\end{equation}
where the two columns of $M$ encode sensitivity to $B_0$ via uniform weights and sensitivity to $g$ via linearly increasing position weights. The sensing matrix is fixed by the field model and the chain geometry and is not a variational degree of freedom.

The phase encoding acts on the probe state as
\begin{equation}
  U(\mathbf{x}) \;=\; \bigotimes_{i=0}^{N-1} \exp\!\left(-\tfrac{i}{2}\,\phi_i\,\sigma_z^{(i)}\right) \;=\; \exp\!\left(-i\,B_0\,G_{B_0} \,-\, i\,g\,G_g\right),
  \label{eq:encoding_unitary}
\end{equation}
with the two generators
\begin{align}
  G_{B_0} \;&=\; \tfrac{1}{2} \sum_{i=0}^{N-1} \sigma_z^{(i)}, \label{eq:gen_B0} \\[2pt]
  G_g     \;&=\; \tfrac{1}{2} \sum_{i=0}^{N-1} (i\,d)\,\sigma_z^{(i)}, \label{eq:gen_g}
\end{align}
the collective $Z$ operator and its position-weighted counterpart. Both generators are diagonal in the computational $\sigma_z$ basis, and their commutator vanishes as an operator identity,
\begin{equation}
  [G_{B_0},\, G_g] \;=\; \tfrac{1}{4} \sum_{i,j} (j\,d) \, [\sigma_z^{(i)}, \sigma_z^{(j)}] \;=\; 0.
  \label{eq:commutator}
\end{equation}
This diagonality is the structural feature that drives the rest of the paper. It removes the generator-noncommutativity obstruction in multiparameter unitary estimation for every probe state, and it reduces the quantum Fisher information matrix to a function of computational-basis probabilities alone, both established in Sec.~\ref{sec:bounds}.

\section{Fisher Information for Joint $(B_0, g)$ Estimation}
\label{sec:fisher}
With the encoder and generators in place, we now define the Fisher information machinery used throughout the rest of the paper. Let $\ket{\psi}$ denote the variationally prepared probe state and
\begin{equation}
  \rho_{\mathbf{x}} \;=\; U(\mathbf{x})\,\ket{\psi}\bra{\psi}\,U^{\dagger}(\mathbf{x})
  \label{eq:encoded_state}
\end{equation}
the encoded state after the phase encoding of Eq.~\eqref{eq:encoding_unitary}. After a single-qubit decoder unitary $V$ (Sec.~\ref{sec:variational}), measurement in the computational basis yields outcome probabilities
\begin{equation}
  p_k(\mathbf{x}) \;=\; \bra{k}\,V\,\rho_{\mathbf{x}}\,V^{\dagger}\,\ket{k},
  \label{eq:outcome_probs}
\end{equation}
indexed by the $2^N$ basis states $\ket{k}$.

The classical Fisher information matrix~\cite{paris_2008} for joint estimation of $\mathbf{x} = (B_0, g)^{\top}$ from the outcomes $\{p_k(\mathbf{x})\}$ is the $2 \times 2$ symmetric matrix
\begin{equation}
  F_{ab} \;=\; \sum_{k}\,\frac{1}{p_k(\mathbf{x})}\,\frac{\partial p_k}{\partial x_a}\,\frac{\partial p_k}{\partial x_b},
  \qquad x_a, x_b \in \{B_0, g\}.
  \label{eq:fim_def}
\end{equation}
The diagonal elements quantify per-parameter sensitivity to $B_0$ and $g$. The off-diagonal element $F_{B_0 g}$, symmetric by construction, has no single-parameter analogue. It measures the statistical correlation between the estimators of $B_0$ and $g$ obtained from the same measurement record and contributes to the joint precision through the Cram\'er-Rao bound $\mathrm{Cov}(\hat{\mathbf{x}}) \succeq F^{-1}$.

The derivatives in Eq.~\eqref{eq:fim_def} are evaluated via the parameter-shift rule for the single-qubit $R_z$ encoders~\cite{schuld_parameter}. Since $p_k$ depends on $\mathbf{x}$ only through the accumulated phases $\{\phi_i\}$ via the linear map $\boldsymbol{\phi} = M\mathbf{x}$ of Eq.~\eqref{eq:sensing_matrix}, the chain rule gives
\begin{equation}
  \frac{\partial p_k}{\partial x_a} \;=\; \sum_{i} M_{ia}\,\frac{\partial p_k}{\partial \phi_i},
\end{equation}
with each elementary derivative recovered from one shifted-circuit pair via
\begin{equation}
  \frac{\partial p_k}{\partial \phi_i} \;=\; \frac{p_k(\phi_i + \pi/2) \,-\, p_k(\phi_i - \pi/2)}{2}.
  \label{eq:param_shift}
\end{equation}
The full $2 \times 2$ FIM is therefore recovered from $2N + 1$ circuit evaluations.

We evaluate the FIM at a fixed operating point $(B_0^{*}, g^{*}) = (0, \pi/100)$, chosen so that the per-spin phases stay in the linear regime of the field expansion (Eq.~\eqref{eq:field_model}) and the parameter-shift evaluation is symmetric across the chain midpoint. The choice fixes only where the local Fisher information is evaluated, not the physical value of $B_0$. The $g^{*} = 0$ point is excluded because the off-diagonal element vanishes there by parity and the FIM becomes diagonal, which does not represent the joint estimation problem at generic points.

The $2 \times 2$ FIM carries three independent elements, and joint estimation quality requires a scalar reduction that weighs all three. The reduction we adopt is the Fisher information determinant~\cite{Liu2020-pk}
\begin{equation}
  \det(F) \;=\; F_{B_0 B_0}\,F_{g g} \,-\, F_{B_0 g}^{2},
  \label{eq:det_fim}
\end{equation}
the standard choice when all parameters are co-equal estimation targets. Geometrically, $\det(F)^{-1/2}$ is proportional to the area of the joint uncertainty ellipse in the $(B_0, g)$ parameter plane at the Cram\'er-Rao level, the region of parameter values indistinguishable from the operating point for a given measurement record. Maximizing $\det(F)$ shrinks this ellipse and improves the joint estimation precision for both parameters simultaneously. A probe state sensitive to both parameters but with statistically correlated estimators produces a large $F_{B_0 g}^{2}$ and therefore a small $\det(F)$, even when both diagonal elements are large. The GHZ state, derived in Sec.~\ref{sec:bounds:ghz}, is the maximally degenerate case of this phenomenon.

\section{Precision Bounds}
\label{sec:bounds}
This section establishes the classical and quantum benchmarks against which the variational results of Sec.~\ref{sec:results} are measured.
\subsection{Generator Commutativity and Measurement Compatibility}
\label{sec:bounds:compat}

A standing concern in multiparameter quantum estimation is whether the quantum Cram\'er-Rao bound is simultaneously saturable for all parameters, since the optimal measurements for different parameters may be incompatible. A standard compatibility condition for pure probe states under unitary encoding is that the generators commute on the state,
\begin{equation}
  \bra{\psi}\,[G_{B_0},\,G_g]\,\ket{\psi} \;=\; 0.
  \label{eq:compat_condition}
\end{equation}
For our generators, both diagonal in the computational basis by Eqs.~\eqref{eq:gen_B0} and~\eqref{eq:gen_g}, the commutator vanishes as an operator identity by Eq.~\eqref{eq:commutator}, so Eq.~\eqref{eq:compat_condition} holds for every probe state.

Ragy~\textit{et al.}~\cite{compatibility_ragy} formulate compatibility conditions for multiparameter quantum metrology. In the present setting, the generator-compatibility condition is automatic by Eq.~\eqref{eq:commutator}, regardless of the probe state. Thus the usual obstruction from noncommuting generators is absent. However, this does not by itself specify a restricted, experimentally feasible measurement that attains the QFIM. In particular, because the encoding is diagonal, direct computational-basis measurement without an appropriate decoder would generally not extract the phase information. The remaining difficulty is therefore concentrated in the joint choice of probe state and decoder/readout: the variational search aims to produce both large diagonal Fisher information elements and a small off-diagonal cross-correlation, yielding a full-rank, high-determinant classical Fisher information matrix.
\subsection{Standard Quantum Limit}
\label{sec:bounds:sql}
For separable probe states, each qubit contributes independently to the Fisher information, and an optimal single-qubit measurement saturates the QFIM at every site. The maximum per-qubit Fisher information for a phase parameter is unity, so the SQL QFIM elements are additive sums over the per-qubit contributions. Using $\partial \phi_i / \partial B_0 = 1$ and $\partial \phi_i / \partial g = i d$ from the sensing matrix of Eq.~\eqref{eq:sensing_matrix}, the diagonal and off-diagonal elements evaluate to
\begin{align}
  Q^{\mathrm{SQL}}_{B_0 B_0} \;&=\; \sum_{i=0}^{N-1} 1 \;=\; N, \label{eq:sql_BB} \\[2pt]
  Q^{\mathrm{SQL}}_{g g} \;&=\; \sum_{i=0}^{N-1} (i d)^{2} \;=\; \frac{d^{2} N (N-1)(2N-1)}{6}, \label{eq:sql_gg} \\[2pt]
  Q^{\mathrm{SQL}}_{B_0 g} \;&=\; \sum_{i=0}^{N-1} (i d) \;=\; \frac{d\,N(N-1)}{2}. \label{eq:sql_Bg}
\end{align}
Setting $d = 1$ in natural units, the SQL Fisher information determinant reduces to the closed form
\begin{align}
  \det(Q^{\mathrm{SQL}}) \;&=\; N \cdot \frac{N(N-1)(2N-1)}{6} \,-\, \left[\frac{N(N-1)}{2}\right]^{2} \notag \\
              \;&=\; \frac{N^{2}(N^{2}-1)}{12},
  \label{eq:det_sql}
\end{align}
which scales as $N^{4}/12$ at large $N$. This is the classical separable benchmark, the largest $\det(F)$ any separable probe state can achieve under any measurement. At $N = 2$ no quantum advantage is achievable, with the SQL coinciding with the quantum bound. From $N \geq 3$ onward, surpassing $\det(Q^{\mathrm{SQL}})$ requires entanglement, with the variational results benchmarked against $\det(Q^*)$ derived in Sec.~\ref{sec:bounds:detq}.

The same SQL elements determine the precision available on $g$ alone when $B_0$ is treated as a nuisance parameter rather than a co-equal estimation target. The Schur complement of the SQL QFIM, $Q^{\mathrm{SQL}}_{gg} - (Q^{\mathrm{SQL}}_{B_0 g})^{2} / Q^{\mathrm{SQL}}_{B_0 B_0} = N(N^{2} - 1)/12$, is the marginal SQL precision on $g$ in this framing, and coincides with the Apellaniz~\textit{et al.}~\cite{gradiometry_toth} bound on gradient precision across all reference states they consider on the equidistant chain. The reason is geometric. The spatial-position covariance that controls quantum enhancement in the Apellaniz framework vanishes identically for the equidistant chain, so the GHZ, Dicke, singlet, and totally polarized reference states all reduce to the same marginal gradient precision $N(N^{2}-1)/12$. The nuisance-parameter framing therefore reports no quantum advantage on this geometry for any standard reference state, while the $\det(F)$ objective adopted here treats $B_0$ and $g$ as co-equal estimation targets and admits quantum advantage on the same geometry.
\subsection{GHZ Collapse}
\label{sec:bounds:ghz}
The GHZ state $\ket{\mathrm{GHZ}_{N}} = (\ket{0\cdots 0} + \ket{1\cdots 1})/\sqrt{2}$ is the natural reference state for entanglement-enhanced single-parameter estimation under uniform encoding. We show that this state gives $\det(Q) = 0$ for the joint $(B_0, g)$ problem at every $N$.
For pure states under unitary encoding, the QFIM elements take the form $Q_{ab} = 4\,\mathrm{Cov}_{\psi}(G_{a}, G_{b})$. On $\ket{\mathrm{GHZ}_{N}}$ the symmetric superposition gives $\langle G_{B_0} \rangle = \langle G_g \rangle = 0$. Direct computation of the second moments using the eigenvalues of the two generators on $\ket{0\cdots 0}$ and $\ket{1\cdots 1}$ then yields
\begin{align}
  Q^{\mathrm{GHZ}}_{B_0 B_0} \;&=\; N^{2}, \label{eq:ghz_BB} \\[2pt]
  Q^{\mathrm{GHZ}}_{g g} \;&=\; \left[\frac{N(N-1)}{2}\right]^{2}, \label{eq:ghz_gg} \\[2pt]
  Q^{\mathrm{GHZ}}_{B_0 g} \;&=\; \frac{N^{2}(N-1)}{2}. \label{eq:ghz_Bg}
\end{align}
The off-diagonal element saturates the Cauchy-Schwarz inequality $Q_{B_0 g}^{2} \leq Q_{B_0 B_0}\,Q_{g g}$ that holds for any positive-semidefinite QFIM,
\begin{equation}
  \begin{split}
    \left(Q^{\mathrm{GHZ}}_{B_0 g}\right)^{2}
      \;&=\; \frac{N^{4}(N-1)^{2}}{4} \\
      \;&=\; N^{2} \cdot \left[\frac{N(N-1)}{2}\right]^{2} \\
      \;&=\; Q^{\mathrm{GHZ}}_{B_0 B_0} \cdot Q^{\mathrm{GHZ}}_{g g},
  \end{split}
  \label{eq:ghz_cs_sat}
\end{equation}
so the GHZ QFIM is rank one and
\begin{equation}
  \det\!\left(Q^{\mathrm{GHZ}}\right) \;=\; 0
  \label{eq:ghz_collapse}
\end{equation}
exactly for all $N$. The collapse reflects the structure of the GHZ state in magnetization space, where both generators act non-trivially only on the two extremal sectors $m = \pm N/2$ and produce perfectly correlated responses to $B_0$ and $g$ variations. We develop this sector-structure observation in Sec.~\ref{sec:results:discussion}.
\subsection{Numerical Benchmark $\det(Q^*)$}
\label{sec:bounds:detq}
The best-found benchmark on $\det(F)$ over all pure $N$-qubit probe states is
\begin{equation}
  \det(Q^*) \;=\; \max_{\ket{\psi}}\,\det\!\left(Q(\ket{\psi})\right),
  \label{eq:detq_def}
\end{equation}
the maximum of the QFIM determinant. No closed-form expression is known on the equidistant linear chain for two non-proportional commuting generators, and we estimate $\det(Q^*)$ numerically.
A reduction simplifies the optimization. Both $G_{B_0}$ and $G_g$ are diagonal in the computational basis, with eigenvalues
\begin{equation}
  \lambda_{k}^{(B_{0})} \;=\; \tfrac{1}{2}\sum_{i}(1 - 2\,b_{i}^{(k)}), \qquad
  \lambda_{k}^{(g)} \;=\; \tfrac{1}{2}\sum_{i}(i\,d)(1 - 2\,b_{i}^{(k)}),
  \label{eq:eigenvalues}
\end{equation}
on basis state $\ket{k}$, where $b_{i}^{(k)} \in \{0, 1\}$ are the bits of $k$. Writing the probe state as $\ket{\psi} = \sum_{k} c_{k} \ket{k}$ and setting $p_{k} = |c_{k}|^{2}$, the QFIM elements reduce to
\begin{equation}
  \begin{split}
    Q_{ab} \;=\; 4\,\Bigg[\,&\sum_{k} p_{k}\,\lambda_{k}^{(a)}\,\lambda_{k}^{(b)} \\
    &-\, \Bigg(\sum_{k} p_{k}\,\lambda_{k}^{(a)}\Bigg)\Bigg(\sum_{k} p_{k}\,\lambda_{k}^{(b)}\Bigg)\Bigg],
  \end{split}
  \label{eq:qfim_simplex}
\end{equation}
which depends only on the probability distribution $\{p_{k}\}$ and not on the relative phases of $\{c_{k}\}$. The maximization in Eq.~\eqref{eq:detq_def} therefore reduces from the $2^{N}$-dimensional complex Hilbert space to the $(2^{N}-1)$-dimensional probability simplex.
The maximization on the simplex is solved by a multi-start L-BFGS-B routine~\cite{Byrd1995} with $50$ random initializations, followed by a differential-evolution~\cite{Storn1997} refinement step. The simplex constraint $\sum_{k} p_{k} = 1$ is handled by a softmax parameterization. The resulting values agree to within $10^{-6}$ across the five best restarts at every $N$, indicating convergence to a local optimum on the simplex. We do not certify global optimality at $N \geq 4$. The reported $\det(Q^*)$ values are therefore lower bounds on the true simplex maximum, and the saturation ratios reported in Sec.~\ref{sec:results} are upper bounds on the true saturation fraction. At $N = 3$ the numerical optimum coincides with an analytical six-state superposition that gives $\det(Q^*) = 10.125$, providing an independent check.
Table~\ref{tab:bounds} collects the SQL benchmark of Eq.~\eqref{eq:det_sql} and the best-found benchmark $\det(Q^*)$ for $N = 2$ to $6$ in natural units. At $N = 2$ the two coincide. From $N = 3$ onward, $\det(Q^*)$ exceeds $\det(Q^{\mathrm{SQL}})$ and the ratio grows monotonically with $N$, reaching a factor of $6.2$ at $N = 6$. The variational results of Sec.~\ref{sec:results} are benchmarked against both the SQL and $\det(Q^*)$.
\begin{table}[t]
  \centering
  \caption{Standard quantum limit and numerical benchmarks on $\det(F)$ for $N = 2$--$6$ in natural units ($d = 1$, $\gamma_{e} t = 1$).}
  \label{tab:bounds}
  \begin{tabular}{ccccc}
    \toprule
    $N$ & $\det(Q^{\mathrm{SQL}})$ & $\log\det(Q^{\mathrm{SQL}})$ & $\det(Q^*)$ & $\log\det(Q^*)$ \\
    \midrule
    2 & $1$    & $0.000$ & $1.0$   & $0.000$ \\
    3 & $6$    & $1.792$ & $10.1$  & $2.314$ \\
    4 & $20$   & $2.996$ & $64.0$  & $4.159$ \\
    5 & $50$   & $3.912$ & $225.3$ & $5.418$ \\
    6 & $105$  & $4.654$ & $650.3$ & $6.477$ \\
    \bottomrule
  \end{tabular}
\end{table}

\section{Variational Framework}
\label{sec:variational}

This section specifies the variational circuit used to prepare the entangled probe state, the four-tier decoder hierarchy used to extract information from it, and the optimization protocol used to train both jointly.

\subsection{Dipolar Entangling Hamiltonian}
\label{sec:variational:hamiltonian}

The entangling dynamics between electron spins are governed by the magnetic dipole-dipole interaction~\cite{Dolde2013}
\begin{equation}
  \hat{H}_{\mathrm{int}} \;=\; \sum_{i<j}\,V_{ij}\,\Big[\,J_{l}\,S_{i}^{z}\,S_{j}^{z} \,+\, J_{s}\,\vec{S}_{i}\cdot\vec{S}_{j}\,\Big],
  \label{eq:H_int}
\end{equation}
with pairwise coupling strength
\begin{equation}
  V_{ij} \;=\; \frac{\mu_{0}\,\gamma_{e}^{2}\,\hbar^{2}}{4\pi}\,\frac{1 - 3\cos^{2}\beta_{ij}}{\lVert\vec{r}_{i} - \vec{r}_{j}\rVert^{3}},
  \label{eq:V_ij}
\end{equation}
where $\beta_{ij}$ is the angle between the inter-spin vector and the bias field. The coupling constants $J_{l} = 3$ and $J_{s} = -1$ correspond to the secular and transverse parts of the dipolar interaction, and $\mu_{0} = 6.7 \times 10^{-4}$ in the simulation units adopted here, following the convention of Ref.~\cite{Zheng2022}. For the linear chain along $\hat{x}$ with bias along $\hat{z}$, the geometry gives $\cos\beta_{ij} = 0$ for all pairs, so
\begin{equation}
  V_{ij} \;\propto\; \frac{1}{|i - j|^{3}\,d^{3}}.
  \label{eq:V_ij_chain}
\end{equation}
The all-to-all character of the dipolar coupling distinguishes this ansatz from local-connectivity variational circuits and maps directly to experimentally accessible NV-array control parameters.

\subsection{Parameterized Circuit Ansatz}
\label{sec:variational:ansatz}

We use the layered dipolar ansatz of Refs.~\cite{Zheng2022, srivastava2024pra}, adapted for the linear chain geometry. The probe state is prepared in two stages. A global $R_{y}(\pi/2)$ rotation acting on $\ket{0}^{\otimes N}$ produces the equal-superposition initial state $\ket{+}^{\otimes N}$. The state is then evolved through $L$ entangling layers of the form
\begin{equation}
  \begin{split}
    \hat{U}^{(\ell)}\!\big(t_{1}^{(\ell)},\,\theta_{2}^{(\ell)},\,t_{3}^{(\ell)}\big) \;=\;& R_{y}(\tfrac{\pi}{2})\,e^{-i\,t_{3}^{(\ell)}\,\hat{H}_{\mathrm{int}}}\,R_{y}(-\tfrac{\pi}{2}) \\
    & \times\,R_{x}\!\big(\theta_{2}^{(\ell)}\,\pi\big)\,e^{-i\,t_{1}^{(\ell)}\,\hat{H}_{\mathrm{int}}},
  \end{split}
  \label{eq:layer_unitary}
\end{equation}
with three trainable parameters $(t_{1}^{(\ell)}, \theta_{2}^{(\ell)}, t_{3}^{(\ell)})$ per layer. The full $L$-layer probe state is
\begin{equation}
  \big|\psi(\vec{\theta})\big\rangle \;=\; \Bigg[\prod_{\ell=1}^{L}\,\hat{U}^{(\ell)}\Bigg]\,\ket{+}^{\otimes N},
  \label{eq:probe_state}
\end{equation}
parameterized by the encoder vector $\vec{\theta} = \{(t_{1}^{(\ell)}, \theta_{2}^{(\ell)}, t_{3}^{(\ell)})\}_{\ell=1}^{L}$ of dimension $3L$. We study circuit depths $L = 1, 2, 3$ in this work.

The dipolar evolution $e^{-i t \hat{H}_{\mathrm{int}}}$ in Eq.~\eqref{eq:layer_unitary} is implemented via first-order Trotterization with $m = 400$ steps per evolution block, using \texttt{qml.ApproxTimeEvolution} in PennyLane~\cite{Bergholm2018-zf}. We adopt the same step count as in Ref.~\cite{srivastava2024pra}, where Fisher information was stable to within $10^{-3}$ relative deviation against finer discretizations.

\subsection{Decoder Hierarchy}
\label{sec:variational:decoder}
After phase encoding, a trainable single-qubit decoder unitary $V$ is applied before the computational-basis measurement. We study four decoder tiers of increasing expressiveness, summarized in Table~\ref{tab:tiers}. The four tiers form a nested hierarchy T1 $\subset$ T2 $\subset$ T3 $\subset$ T4, so the achievable $\det(F)$ at each tier is monotonically non-decreasing with tier index. T1 is the fixed Ramsey readout with no free decoder parameters. T2 generalizes T1 to a global trainable rotation, T3 separates qubit $0$ from the rest because its position $x_0 = 0$ gives $\partial \phi_0 / \partial g = 0$ and contributes only to $F_{B_0 B_0}$, and T4 is the most expressive single-qubit decoder under $Z$-basis measurement.

\begin{table}[htbp]
  \centering
  \caption{Four decoder tiers studied in this work}
  \label{tab:tiers}
  \begin{tabular}{lll}
    \toprule
    Tier & Decoder structure & Parameter count \\
    \midrule
    T1 & Fixed Ramsey, $R_{x}(\pi/2)$ on all qubits & $0$ \\
    T2 & Shared $R_{z}(\alpha)\,R_{x}(\beta)$ on all qubits & $2$ \\
    T3 & Independent rotation on $q_{0}$, shared on rest & $4$ \\
    T4 & Independent $R_{z}\,R_{x}$ per qubit & $2N$ \\
    \bottomrule
  \end{tabular}
\end{table}
\subsection{Optimization Objective and Protocol}
\label{sec:variational:protocol}
The joint optimization variable is the concatenation of the encoder parameters $\vec{\theta}$ and the decoder parameters of the chosen tier, optimized in a single CMA-ES run. The optimization objective is the regularized log-determinant of the Fisher information matrix,
\begin{equation}
  \mathcal{L}(\vec{\theta}) \;=\; \log\det\!\left(F(\vec{\theta}) + \lambda\,\mathbb{I}_{2}\right),
  \label{eq:objective}
\end{equation}
with $\lambda = 10^{-6}$. Maximization of $\mathcal{L}$ is equivalent to minimization of the joint uncertainty ellipse area at the Cram\'er-Rao level (Sec.~\ref{sec:fisher}). The regularization prevents $\log\det$ from diverging when the optimizer transiently visits near-rank-deficient FIMs and is negligible compared to $F$ at every converged probe state reported here.

Fitness evaluation uses the parameter-shift rule of Sec.~\ref{sec:fisher}, requiring $2N + 1$ circuit calls per evaluation and returning all three independent FIM elements from a single batch of shifted circuits. The optimizer is the Covariance Matrix Adaptation Evolution Strategy~\cite{hansen2006}, a gradient-free evolutionary algorithm that adapts a multivariate Gaussian search distribution from rankings of sampled candidates. The optimization vector contains at most $3L + 2N = 21$ parameters (Tier~4, $L=3, N=6$), and the landscape is multimodal at large $N$. CMA-ES traverses such landscapes more reliably than local gradient-based methods at this scale, and avoids the quadratic cost of nested parameter-shift gradients.

To accelerate convergence at deeper circuits, we employ a layerwise warm-start across $L$~\cite{Skolik2021}. Depth $L = 1$ is optimized from a broad random initial distribution. The best parameters at depth $L$ are then used to seed depth $L + 1$ by appending a new layer initialized near zero, leaving the previously learned parameters intact. This preserves correlations learned at shallower depths and limits the search at each new $L$ to the newly introduced degrees of freedom.

To distinguish converged optima from seed-dependent artifacts, every $(L, N, \mathrm{tier})$ cell is run from at least the standard CMA-ES seed set $\{204, 604, 1204, 2004, 3004\}$, with additional seeds added at harder configurations until the best-seed value stops changing across restarts. The reported $\det(F)$ at each cell is the maximum across all seeds, and the per-seed spread is reported in Appendix~\ref{app:seed_variance}.

\section{Results and Discussion}
\label{sec:results}

\subsection{Scaling and Saturation of the Quantum Bound}
\label{sec:results:scaling}
\begin{figure*}[htbp]
    \centering
    \includegraphics[width=\textwidth]{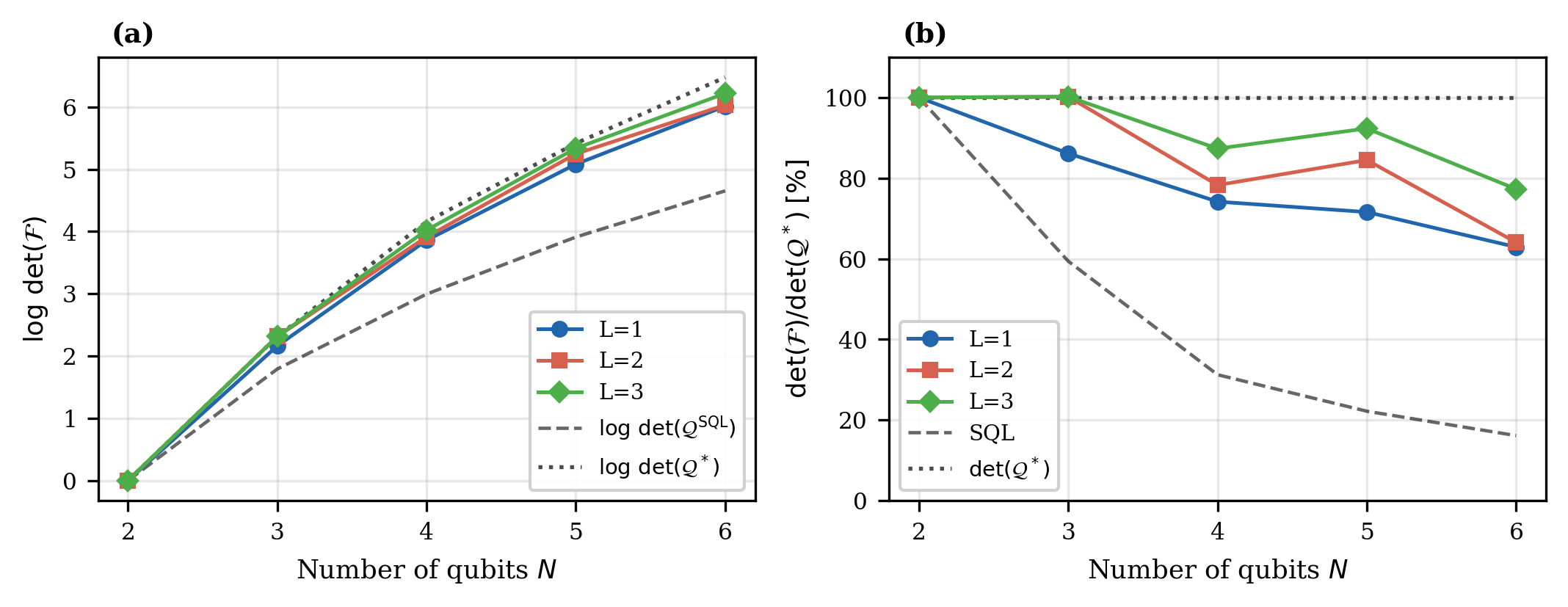}
    \caption{%
    Scaling of the joint Fisher information determinant under fixed Ramsey readout (Tier~1) for circuit depths $L = 1, 2, 3$.
    (a)~$\log\det(F)$ as a function of qubit number $N$, with the SQL benchmark $\log\det(Q^{\mathrm{SQL}})$ (dashed) and the numerical bound $\log\det(Q^*)$ (dotted) shown for reference.
    (b)~The same data normalized by the found benchmark, $\det(F)/\det(Q^*)$, expressed as a percentage on a linear scale.%
    }
    \label{fig:scaling_main}
\end{figure*}

Having specified the encoder, the decoder hierarchy, and the optimization protocol, we now report the variational performance across the optimization grid. Figure~\ref{fig:scaling_main} shows the optimized $\det(F)$ as a function of qubit number $N$ for circuit depths $L = 1, 2, 3$ under fixed Ramsey readout (Tier~1, Sec.~\ref{sec:variational:decoder}). Panel~(a) reports $\log\det(F)$ on the same axes as the SQL benchmark $\log\det(Q^{\mathrm{SQL}})$ and the numerical bound $\log\det(Q^*)$ obtained by full-simplex optimization (Sec.~\ref{sec:bounds:detq}). Panel~(b) re-expresses the same data as a percentage of the best-found benchmark, $\det(F)/\det(Q^*)$, making the saturation level directly visible.

Throughout this section we report Tier~1 results in the main text, with the full four-tier comparison deferred to Appendix~\ref{app:tier_comparison}. Tier~1 corresponds to the fixed Ramsey readout with no trainable decoder parameters, so the saturation values reported here measure the encoder alone against the joint-estimation benchmark, isolating the encoder's contribution from any measurement-basis tuning. The saturation ratios should themselves be read as upper bounds on the true saturation fraction, since $\det(Q^*)$ is obtained by numerical optimization on the simplex and is a lower bound on the true maximum at $N \geq 4$ (Sec.~\ref{sec:bounds:detq}).

At $N = 2$ the SQL coincides with $\det(Q^*)$, so no quantum advantage is achievable, and the optimizer recovers the separable bound. From $N = 3$ onward, the gap between SQL and $\det(Q^*)$ opens rapidly, and the variational curves track $\det(Q^*)$ closely at small system sizes, exceeding the SQL at every cell of the grid. At $L = 3$ the optimizer reaches $\det(F)/\det(Q^*) = 1.00$ at $N = 3$, $0.87$ at $N = 4$, $0.92$ at $N = 5$, and $0.77$ at $N = 6$, with the absolute advantage over the SQL reaching a factor of roughly $4.2$ at $N = 5$.

The approach to $\det(Q^*)$ tightens with depth at every $N$, and the largest residual gaps occur at the largest system sizes for the shallowest circuits, consistent with the expressivity-mismatch interpretation developed in Sec.~\ref{sec:results:discussion} below. The seed-variance analysis in Appendix~\ref{app:seed_variance} confirms that the best-seed values reported here are stable across independent CMA-ES runs, with the residual variability concentrated at the larger $N$. The percentage curves in panel~(b) decrease with $N$ at every $L$ even though the absolute $\det(F)$ grows steeply with $N$, because $\det(Q^*)$ grows faster with $N$ than the variational $\det(F)$ does.

\subsection{Structure of the Optimal Probe State}
\label{sec:results:state_structure}

\begin{figure*}[htbp]
    \centering
    \includegraphics[width=\textwidth]{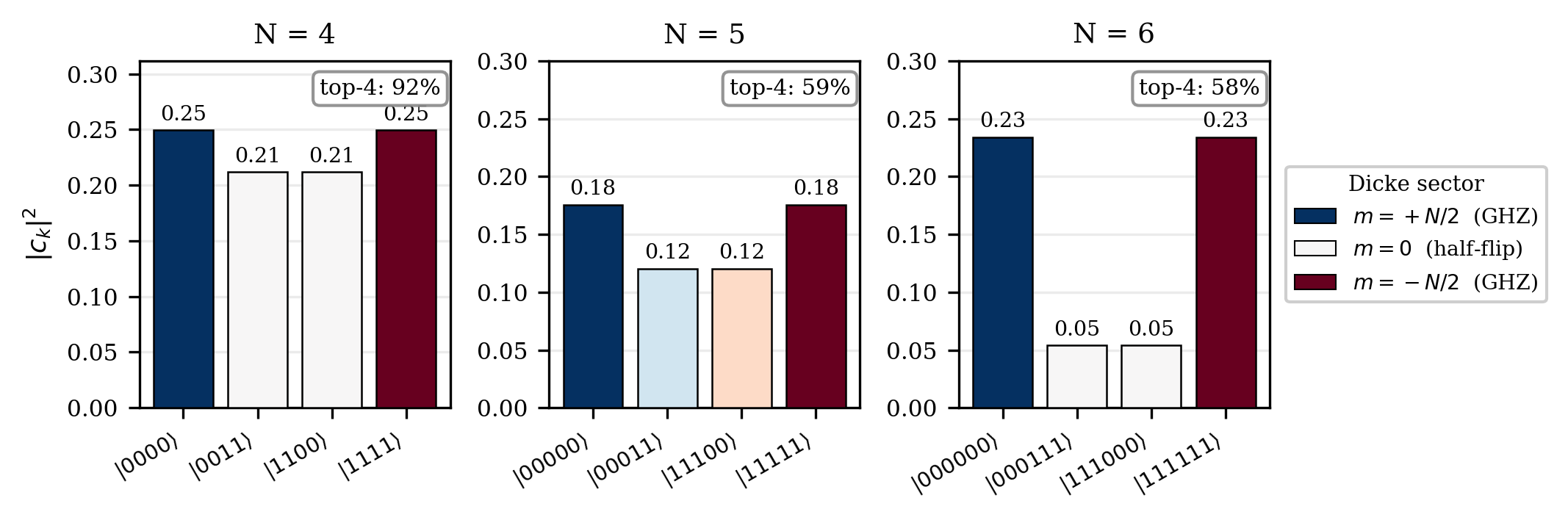}
    \caption{%
        Computational-basis structure of the variationally optimized probe state at $L = 3$, Tier~1, for $N = 4, 5, 6$. Each panel shows the probabilities $|c_k|^2$ of the four largest-weight basis states, with the cumulative top-four weight annotated at the top right. Bars are colored by Dicke sector $m = \tfrac{1}{2}\sum_i (1 - 2 b_i)$. The half-flip pair occupies the $m = 0$ sector at even $N$ and the $m = \pm 1/2$ sectors at odd $N$, accounting for the color variation across panels.%
    }
    \label{fig:state_structure}
\end{figure*}

Figure~\ref{fig:state_structure} reports the four largest computational-basis amplitudes of the optimized state at $L = 3$, Tier~1, for $N = 4, 5, 6$. The same four basis strings appear at the top of the distribution at every $N$, namely the two GHZ extrema $|0 \cdots 0\rangle$ and $|1 \cdots 1\rangle$, and the two half-chain-flip strings $|0\cdots 0\, 1 \cdots 1\rangle$ and $|1 \cdots 1\, 0 \cdots 0\rangle$. We refer to this set of four strings as the GHZ + half-flip motif. The motif identity is not specific to $L = 3$ or to Tier~1. Across all $60$ optimized cells of the grid, these four strings are the four largest-weight basis states without exception.

The motif weights obey a regular structure across the grid. The two GHZ extrema carry equal weight, the two half-flip strings carry equal weight, and the GHZ pair is weighted at least as heavily as the half-flip pair at every cell. Reading off Fig.~\ref{fig:state_structure}, the cumulative motif weight drops sharply from $0.92$ at $N = 4$ to $0.59$ at $N = 5$, and remains near that level at $N = 6$ ($0.58$). The residual weight outside the motif grows correspondingly, from $0.08$ at $N = 4$ to roughly $0.42$ at $N = 5, 6$.

The cumulative motif weight is therefore neither monotonically falling with $N$ nor faithfully tracking the saturation values of Fig.~\ref{fig:scaling_main}. The $N = 4$ cell has the highest motif concentration and the lowest saturation among the three, while the $N = 5$ cell has substantially lower motif concentration and the highest saturation. The relationship between motif weight and saturation is therefore the puzzle that the discussion below resolves.

Despite the motif containing the GHZ extrema, the optimized state is itself far from a GHZ state at every system size. The squared overlap $|\langle \mathrm{GHZ}_N | \psi(\vec{\theta}) \rangle|^2$ remains between $0.35$ and $0.55$ for $N = 3$--$6$ across all depths, bounded above by the GHZ-pair weight in the variational state. On the same dipolar ansatz under uniform single-parameter encoding~\cite{srivastava2024pra}, by comparison, the optimizer drives the probe state toward the GHZ manifold with circuit depth and reaches GHZ fidelity above $0.95$ at $L = 4$.

\subsection{Discussion}
\label{sec:results:discussion}

We now interpret the results of Sec.~\ref{sec:results:scaling} and Sec.~\ref{sec:results:state_structure} jointly. The connection between saturation behavior and motif structure follows from the objective, the generator structure, and the limited expressivity of a fixed-depth encoder, in that order.

The optimizer's avoidance of the GHZ manifold, evident from the moderate fidelities reported in Sec.~\ref{sec:results:state_structure}, is a direct consequence of the objective. The $\log\det(F)$ landscape penalizes any state that saturates the Cauchy-Schwarz inequality on $F$, and the GHZ state does so identically (Sec.~\ref{sec:bounds:ghz}). The variational fidelities are therefore the empirical counterpart of the analytical GHZ collapse, with no external constraint imposed on the optimizer.

The Dicke-sector structure of the motif explains why these four basis strings emerge as the optimal subspace. The magnetization quantum number $m = (N - 2k)/2$, with $k$ the Hamming weight of a basis string, partitions the $2^N$-dimensional Hilbert space into sectors, and the two encoding generators couple to this partition in complementary ways. The collective generator $G_{B_0} = \tfrac{1}{2} \sum_i \sigma_z^{(i)}$ acts as the total magnetization, with eigenvalues constant within each sector, so generating variance in $G_{B_0}$ requires amplitude spread across distinct sectors. The position-weighted generator $G_g = \tfrac{1}{2} \sum_i (i d) \, \sigma_z^{(i)}$ distinguishes basis strings of equal magnetization but different spatial arrangement, so it admits a variance contribution from amplitude spread within a single sector.

The motif is the minimal structure that satisfies both requirements. The two GHZ extrema occupy the extremal sectors $m = \pm N/2$ with Hamming weights $k = 0, N$, supplying inter-sector spread for $G_{B_0}$. The half-flip pair occupies a near-central pair of sectors, $m = 0$ at even $N$ with $k = N/2$, and $m = \pm 1/2$ at odd $N = 5$ with $k = 2, 3$ falling in adjacent sectors because $N$ is odd. This pair supplies intra-sector spread for $G_g$ at a magnetization that does not displace $\langle G_{B_0} \rangle$ from zero. Removing either pair collapses the construction. Without the GHZ extrema, the state lives in a single near-central sector and $F_{B_0 B_0} = 0$. Without the half-flip pair, the state reduces to the GHZ state, and the rank-one collapse of Sec.~\ref{sec:bounds:ghz} returns.

The sector argument identifies the motif as the dominant subspace at every $N$, but it does not by itself fix the relationship between motif weight and saturation. Saturation depends on alignment between the variational distribution $\{|c_k^{\mathrm{var}}|^2\}$ and the simplex-optimal distribution $\{|c_k^*|^2\}$ across the entire computational basis, while the motif weight reports only the variational distribution restricted to the four motif strings. The two coincide as a performance proxy when the simplex-optimal distribution itself has support only on the motif.

At $N = 4$ this condition holds. The simplex optimum is the equal superposition $(|0000\rangle + |1111\rangle + |0011\rangle + |1100\rangle)/2$, with $|c_k^*|^2 = 0.25$ on the motif strings and zero elsewhere, and the variational saturation deficit is bounded above by the variational off-motif weight, which at $N = 4$ is $0.08$. At $N \geq 5$ the simplex optimum itself spreads beyond the motif, and the variational state also spreads, with $0.42$ of its weight outside the motif at $N = 5, 6$. Off-motif weight is no longer wasted, and the saturation depends on how well the variational off-motif distribution aligns with the benchmark's off-motif distribution. This is why the $N = 5$ cell achieves higher saturation than $N = 4$ despite carrying lower motif weight, and why the $N = 6$ cell achieves lower saturation than $N = 5$ despite comparable motif weight. The variational off-motif distribution at $N = 6$ aligns less well with the benchmark than the corresponding distribution at $N = 5$.

The saturation deficit at large $N$ reflects an expressivity mismatch rather than an optimization failure. The encoder has $3L$ trainable parameters at depth $L$, fixed across the grid, while the dimension of the simplex-optimal support grows with $N$. At small $N$ the support is concentrated on the motif strings, and three entangling layers prepare a state aligned with it. At $N = 6$ the support is substantially larger than the motif, and three layers can place amplitude outside the motif but not in the directions the spread benchmark requires. The same three-layer ansatz, therefore, cannot reach the same fraction of the optimum even with fully converged optimization, consistent with the seed-variance evidence in Appendix~\ref{app:seed_variance} that the best-seed values are stable across independent restarts.

\section{Conclusion and Outlook}
\label{sec:conclusion}

We have presented a variational framework for joint estimation of a uniform magnetic field $B_{0}$ and its spatial gradient $g$ on a linear chain of dipolar-interacting spin-1/2 systems, with $\det(F)$ as the optimization objective and a layered dipolar circuit as the variational ansatz. The framework treats the sensing protocol as a constrained hardware-design problem in which the geometry fixes the two encoding generators, the dipolar interaction supplies the native entangling resource, the circuit depth is capped to reflect near-term implementation constraints, and the decoder hierarchy quantifies the value of additional measurement-basis control.

Two analytical results frame the task. The two encoding generators commute as an operator identity because both are diagonal in the computational basis, so the quantum Cram\'er-Rao bound is simultaneously saturable on every probe state, and the difficulty is concentrated in the choice of probe rather than in measurement incompatibility. The GHZ state, the standard reference for entanglement-enhanced single-parameter sensing, gives a rank-one QFIM with $\det(Q^{\mathrm{GHZ}}) = 0$ for every $N$. No closed-form optimal probe is known on the equidistant chain, motivating the variational framework adopted here.

The numerical results show that the layered dipolar ansatz tracks the best-found benchmark closely at small system sizes and falls off at the largest sizes studied, in a way that the seed-variance analysis attributes to ansatz expressivity rather than optimization failure. The optimized probe state at every cell concentrates on a four-string motif of the two GHZ extrema and two half-chain-flip strings, which emerges from the Dicke-sector structure of the two generators as the minimal subspace that feeds both diagonal QFIM elements without saturating the Cauchy-Schwarz inequality. The diagonal generator structure also makes the classical Fisher information depend only on basis-state probabilities, which predicts that a trainable single-qubit decoder before computational-basis measurement should add little over fixed Ramsey readout. The joint CMA-ES optimization confirms this prediction empirically across the four-tier decoder hierarchy, with gains above Tier 1 bounded by a few percentage points at $L = 3$ on every cell of the grid.

The motivating context is wide-field NV-array magnetometry, where the bias field drifts on integration timescales, and the present framework removes the calibration floor that a nuisance-parameter treatment imposes. Concrete settings include optical magnetic detection of single-neuron action potentials~\cite{barry2016neural}, magnetic imaging of living cells and bacterial samples~\cite{Lesage2013}, and micrometer-scale magnetic imaging of geological specimens with the quantum diamond microscope~\cite{glenn2016geological}. The same structure applies to trapped-ion chains and Rydberg-atom arrays under analogous joint-encoding conditions.

The proof-of-principle scope is deliberate. The four-string motif places substantial weight on the two GHZ extrema, and these components are precisely the ones expected to dephase catastrophically on the readout basis under realistic decoherence. Quantifying this trade-off, together with extensions to non-equidistant geometries, ancilla-assisted protocols, and noise-aware variational training, is the subject of ongoing work.

%
%
\section*{Acknowledgment}

KPS thanks the U.S. Department of Energy, Office of Science, Advanced Scientific Computing Research (ASCR) program, for support under Award Number DE-SC0026264, and PQI Community Collaboration Awards. GD also acknowledges support from NSF Award No. 2304998. This research was also supported in part by the University of Pittsburgh Center for Research Computing and Data (RRID: SCR\_022735) through the computational resources provided. The work utilized the HTC and H2P clusters, which are supported by NIH award number S10OD028483 and NSF award number OAC-2117681. The authors used Anthropic's Claude AI model for language refinement and presentation improvement of the manuscript. The authors take full responsibility for the final content.
\bibliographystyle{IEEEtran}
\bibliography{sections/references}
\appendices


%
\section{Seed Variance}
\label{app:seed_variance}

The CMA-ES landscape for $\log\det(F)$ is non-convex at the larger system sizes studied here. This appendix reports the per-seed spread across the optimization grid and shows that the best-seed values used in the main text are stable across independent restarts.

\begin{figure}[htbp]
    \centering
    \includegraphics[width=\columnwidth]{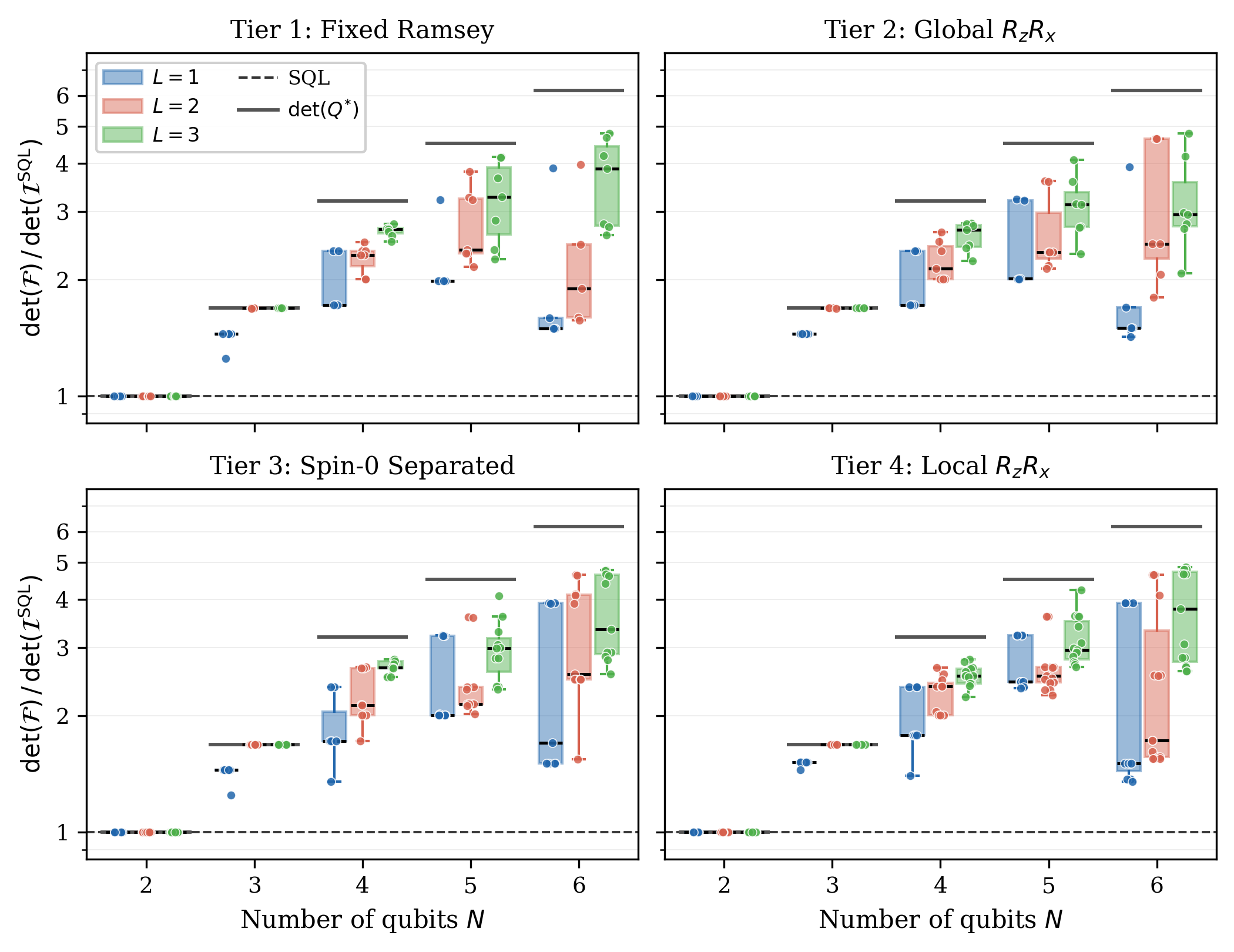}
    \caption{%
        Per-seed values of $\det(F) / \det(Q^{\mathrm{SQL}})$ across the four decoder tiers, grouped by $N$ and color-coded by depth $L$. Each box reports the inter-quartile range across all CMA-ES seeds at that $(L, N, \text{tier})$ cell, with whiskers and individual seed outcomes overlaid. The dashed line marks the SQL; the solid horizontal segments above each $N$ mark the best-found benchmark $\det(Q^*) / \det(Q^{\mathrm{SQL}})$.%
    }
    \label{fig:seed_variance}
\end{figure}

The seed spread grows with $N$ at fixed $L$ and broadly shrinks with depth. For Tier~1 at $L = 3$, the seed spread of $\log\det(F)$ is below $1\%$ at $N = 2, 3$ and rises to $11.5\%$ at $N = 5$ and $9.8\%$ at $N = 6$, with the worst values in the grid ($15.9$--$17.7\%$) at $L = 1, N = 6$. The pattern reflects the increasing dimensionality of the optimization landscape. The encoder parameter count is fixed at $3L$ while the dimension of the simplex-optimal support grows with $N$ (Sec.~\ref{sec:results:discussion}), so the probability that any individual seed converges to a sub-optimal basin grows accordingly.

The reported best-seed values are stable across independent restarts. For every cell at $L = 2$ and $L = 3$, the best seed was recovered at multiple restarts within a tolerance of $0.5\%$ of the reported $\log\det(F)$, indicating a converged local optimum rather than a single statistical fluctuation. The $L = 3$ cells with the largest residual spread, namely $N = 5$ and $N = 6$, are precisely the cells where the saturation deficit reported in Sec.~\ref{sec:results:scaling} is largest. The two are not independent. The same expressivity mismatch that prevents the variational state from fully populating the simplex-optimal support at large $N$ also makes the optimization landscape harder to traverse, and additional seeds at these cells do not appreciably improve the best-seed value. The residual gap to $\det(Q^*)$ at large $N$ therefore reflects an ansatz-expressivity limitation rather than a finite-seed-count artifact.

\section{Decoder Tier Comparison}
\label{app:tier_comparison}

The main text reports Tier~1 (fixed Ramsey) results throughout. This appendix justifies that choice by quantifying decoder expressivity across the grid and connecting the empirical near-tier-agreement to the diagonal-generator reduction of Sec.~\ref{sec:bounds:detq}.

\begin{figure}[htbp]
    \centering
    \includegraphics[width=\columnwidth]{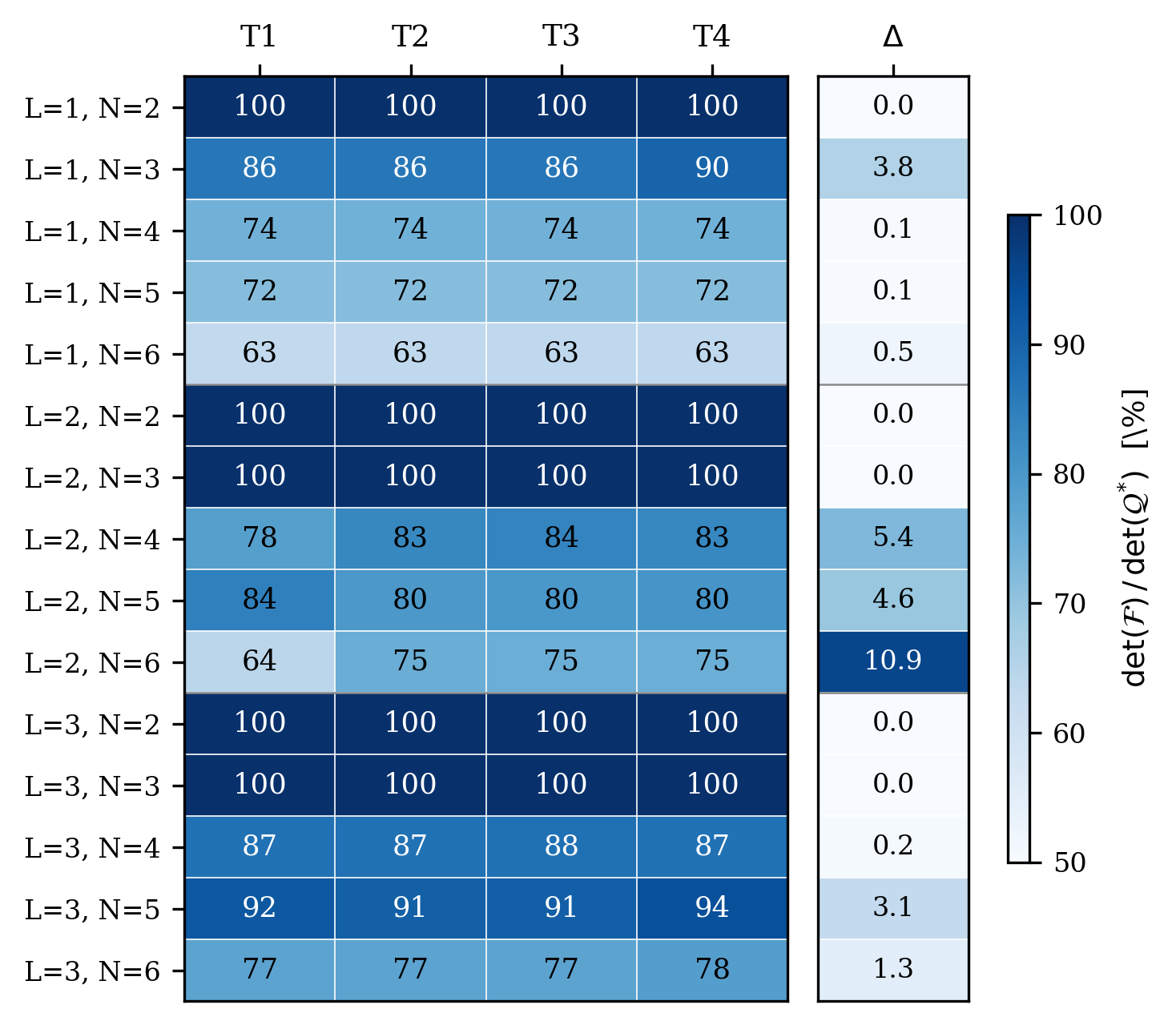}
    \caption{%
        Saturation of the best-found benchmark $\det(F)/\det(Q^*)$ across the four decoder tiers and all $(L, N)$ cells. The rightmost column reports the spread $\Delta$ between the best and worst tier in each row, in percentage points of the benchmark.%
    }
    \label{fig:tier_heatmap}
\end{figure}

Figure~\ref{fig:tier_heatmap} reports $\det(F)/\det(Q^*)$ for all four decoder tiers across the full $(L, N)$ grid. The four tiers agree to within $5.5$ percentage points of the benchmark at $14$ of $15$ cells, and the maximum spread at $L = 3$ is $3.1$ percentage points (at $N = 5$). The exception is $(L = 2, N = 6)$, where Tier~1 reaches only $64\%$ of the benchmark while Tiers~2--4 reach approximately $75\%$.

The trend across depth decomposes the gap to the benchmark into encoder and decoder contributions. At $L = 1$, the encoder is shallow and far from the benchmark regardless of measurement basis. The residual encoder gap is $25$--$37$ percentage points at $N \geq 4$, while decoder gains stay below $1$ percentage point. At $L = 2$, the decoder contribution reaches its maximum of $10.9$ percentage points at $N = 6$, but the encoder gap remains the larger contribution for $N \geq 4$. At $L = 3$, the decoder contribution collapses to at most $1.6$ percentage points across the entire grid, with the residual encoder gap of $6$--$22$ percentage points accounting for almost the entire distance to the benchmark.

The near-tier-agreement is the empirical signature of the diagonal-generator reduction of Sec.~\ref{sec:bounds:detq}. The same argument that makes the QFIM depend only on the basis-state probabilities $\{p_k\}$ applies to the FIM under any computational-basis measurement, since the FIM elements involve the same probability gradients. A trainable single-qubit decoder before $Z$-basis measurement only reshuffles which probability distribution the FIM is computed over, so once the encoder produces a state whose basis-state probabilities are well-matched to the joint $(B_0, g)$ problem, the fixed Ramsey readout extracts essentially all available information. The $(L = 2, N = 6)$ exception is consistent with this picture: the encoder there has not yet reached such a distribution, so the additional decoder parameters of Tiers~2--4 pick up some of the slack the encoder leaves behind.

\end{document}